\documentclass[aps,prl,showpacs,twocolumn,superscriptaddress]{revtex4-1}

\usepackage{hyperref,xcolor}
\usepackage{graphicx}
\usepackage{amsmath,amssymb,latexsym,amsfonts,subfigure,bbm}
\usepackage{dsfont,relsize}
\usepackage{color}
\usepackage[utf8]{inputenc}

\newcommand{\psreq}{\vec{\psi}\!\left(\omega\right)}
\newcommand{\psheq}{\hat{\psi}\!\left(\omega\right)}

\newcommand{\treq}{t \!\left(\omega\right)}

\newcommand{\phat}{\hat{\phi}}

\newcommand{\rhoreq}{\rho \! \left( \omega,  \omega_0 \right)}
\newcommand{\coreq}{C \! \left( \omega, \omega_0 \right)}

\newcommand{\wheq}{W \! \left( \omega \right)}

\usepackage{cmap}
\definecolor{rltred}{rgb}{0.75,0,0}
\definecolor{rltgreen}{rgb}{0,0.5,0}
\definecolor{rltblue}{rgb}{0,0,0.75}
\hypersetup{colorlinks,%
hypertexnames=true,%
pdfauthor={xxx},%
pdftitle={xxx},%
pdfview=FitH,%
pdfstartview=FitH,%
pdfpagemode=UseNone,%
bookmarksopen=true,%
bookmarksnumbered=true,%
pdfhighlight=/I,%
linkcolor=rltred,%
citecolor=rltgreen,%
linktocpage=true}

\begin{document}
\date{April 3, 2017}
\setlength{\tabcolsep}{1pt}
\title{Super- and Anti-Principal Modes in Multi-Mode Waveguides}

\author{Philipp Ambichl}
\affiliation{Institute for Theoretical Physics,
Vienna University of Technology (TU Wien), A-1040, Vienna, Austria, EU}
\author{Wen Xiong}
\affiliation{Department of Applied Physics,
Yale University, New Haven, Connecticut 06520, USA}
\author{Yaron Bromberg}
\affiliation{Department of Applied Physics,
Yale University, New Haven, Connecticut 06520, USA}
\author{Brandon Redding}
\affiliation{Department of Applied Physics,
Yale University, New Haven, Connecticut 06520, USA}
\author{Hui Cao}
\affiliation{Department of Applied Physics,
Yale University, New Haven, Connecticut 06520, USA}
\author{Stefan Rotter}
\affiliation{Institute for Theoretical Physics,
Vienna University of Technology (TU Wien), A-1040, Vienna, Austria, EU}

\begin{abstract}
We introduce a new type of states for light in multimode waveguides featuring strongly enhanced or reduced spectral correlations. Based on the experimentally measured multi-spectral transmission matrix of a multimode fiber, we generate a set of states that outperform the established ``principal modes'' in terms of the spectral stability of their output spatial field profiles. Inverting this concept also allows us to create states with a minimal spectral correlation width, whose output profiles are considerably more sensitive to a frequency change than typical input wavefronts. The resulting ``super-'' and ``anti-principal'' modes are made orthogonal to each other even in the presence of mode-dependent loss. By decomposing them in the principal mode basis, we show that the super-principal modes are formed via interference of principal modes with closeby delay times, whereas the anti-principal modes are a superposition of principal modes with the most different delay times available in the fiber. Such novel states are expected to have broad applications in fiber communication, imaging, and spectroscopy.     
\end{abstract}
\maketitle

\section{I. Introduction}

Optical fibers are the enabling tools for transferring the ever-increasing data volumes characteristic for our information age. 
Since single-mode fibers will soon reach their capacity limit \cite{richardson_space-division_2013,Essiambre2010}, the next step will be to switch to multi-mode fibers, whose spatial or transverse degree of freedom can be exploited for communication \cite{kahn_communications_2017,randel20116,ryf2012mode,ryf2011space,ryf201332,fontaine201530,salsi2011transmission,carpenter2012degenerate,carpenter2012all} and imaging applications \cite{flusberg2005fiber,ChoiPRL12,ploschner2015seeing,vcivzmar2012exploiting,ploschner2015compact,cizmar_shaping_2011,Amitonova16,papadopoulos2013high,morales2015two,stasio2016calibration,loterie2017bend,kolenderska2015scanning}. While many transverse modes in a fiber hold promise for a considerable increase in the information transmission capacity, the crucial challenge is to overcome the detrimental effects of random mode mixing, induced by fiber imperfections and external perturbations \cite{raddatz1998influence,patel2002enhanced,shen2005compensation,Panicker2008,Panicker2009,morales2015delivery,parto2016systematic,carpenter2016complete}. In combination with the unavoidable modal dispersion, this leads to spatial and temporal distortion of optical signals, especially when the mode mixing is strong in the fiber. 

A way out of this predicament is provided by the eigenstates of the Wigner-Smith time delay operator \cite{wigner1955lower,smith1960lifetime,Fan2005,Juarez2012,Shemirani2009,hokahn2011,Rotter2011,franzbulow2012,milione2015determining,Carpenter2015,Xiong2015,XiongOE17,carpenter2016comparison}, also called the ``principal modes" (PMs), which were introduced to suppress modal dispersion, allowing optical pulses to be sent through a multimode fiber (MMF) without distortion. In addition, the PMs maintain the spatial coherence so that the output field can be focused to a diffraction-limited spot or converted to any spatial pattern \cite{Carpenter2015, Xiong2015}. The ability of maintaining temporal pulse shape and spatial coherence of a broadband signal amid strong spatio-temporal scattering is desirable for many applications including fiber communication, microscopy, MMF lasers and amplifiers \cite{kim2012maximal,papadopoulos2012focusing,aulbach2011control,richardson2010high,jauregui2013high,florentin2017shaping}. Moreover, the PMs have great potential for quantum information protocols as they can be used to synthesize decoherence-free high-dimensional quantum states \cite{peruzzo2010multimode,peruzzo2010quantum,defienne2016two}. Thanks to the recent advances in wavefront shaping techniques \cite{vellekoop2007focusing,vellekoopNatPhoton10,Popoff2010,Popoff2011,Popoff2010a,katz2011focusing,Mosk2012,rotter2017light}, the above desirable features of PMs are now also accessible in experiments using MMFs with weak or strong mode coupling \cite{Carpenter2015, Xiong2015,XiongOE17}. 

In spite of their promising potential, PMs suffer from two major limitations. First, they have a finite spectral width, limiting the bandwidth of input signals that can maintain the temporal pulse shape and spatial coherence after transmitting through the MMF. This limitation is severe for MMFs with strong mode mixing, where the bandwidth of PMs is particularly narrow. Secondly, PMs are non-orthogonal to each other in a MMF with mode-dependent loss (MDL), resulting in crosstalk between them. A fundamental question is thus whether an orthogonal set of states exist that outperform PMs in terms of the spectral correlation width, especially in the strong mode coupling regime where a gain in bandwidth is in the highest demand. If such special states indeed exist, a practical question is how to generate them experimentally - an operational procedure which could have a broad impact on fiber applications in communication, imaging, nonlinear microscopy, laser amplifiers, and quantum technology.      

In a seemingly unrelated context, it has recently been shown that an individual multimode fiber can function as a spectrometer with ultrahigh resolution, broad bandwidth and low loss \cite{redding_using_2012,redding_all-fiber_2013,redding2014noise,redding2014high,redding2016evanescently,liew2016broadband,coluccelli2016optical,valley2016multimode}. The output speckle pattern of this MMF is formed by interference of the guided fiber modes, and it changes with input frequency. The working principle of the fiber spectrometer is now to use the frequency-dependent speckle pattern as a fingerprint to recover the input spectrum.  To further increase the resolving power of this device, one needs to enhance the frequency sensitivity of its output speckle patterns. The key question here is whether it is possible to achieve this by creating a special input state whose output field pattern is considerably more sensitive to a frequency change than that of a typical input to the MMF. The crucial point is, in other words, whether one can use the spatial degree of freedom at the input to accelerate the spectral decorrelation at the output of a MMF.

A practical application of MMFs in biomedical imaging is to reduce the spatial coherence of a broadband light source for parallel optical coherence tomography (OCT) \cite{dhalla2010crosstalk}. When imaging through turbid media such as biological tissue, the suppression of spatial coherence of the illuminating light prevents resolution loss from crosstalk due to coherent multiple scattering. If it was possible to create a state in the MMF with much reduced spectral correlation, one could greatly suppress the spatial coherence at the MMF output by launching broadband light into such a state.

Here we introduce such novel states of light that have the aforementioned unique characteristics and we generate them experimentally by addressing the spatial degrees of freedom of a MMF with wavefront shaping techniques. As we demonstrate explicitly, exceeding the already efficient performance of PMs is possible by making use of the information stored in the multi-spectral transmission matrix of a MMF. Based on an optimization procedure operating on such a matrix \cite{andreoli_deterministic_2015}, we are able to create a set of ``super-PMs'' that not only have a significantly increased spectral stability as compared to the most stable PM available in the same fiber, but that also have the great advantage of being mutually orthogonal even in the presence of MDL. Moreover, our optimization procedure can be applied to generate ``anti-PMs'' featuring an extremely narrow spectral correlation width -- a property that holds promise for spectroscopy and sensing applications with MMFs. 

Intuitively, the physical mechanisms for the formation of both ``super-PMs'' and ``anti-PMs'' can be understood in the basis of PMs. In the absence of MDL, ``super-PMs'' are formed by PMs with similar delay times and the interference effect between them brings about a broader bandwidth for the ``super-PMs''. On the contrary,  ``anti-PMs'' are composed of PMs with extremely different delay times, which directly results in a bandwidth even narrower than that of random inputs.  More importantly, our analysis illustrates that the PMs provide a powerful basis for synthesizing new types of states with unique spatial, temporal and spectral characteristic.

%This article is organized as follows. In section II, we recapture the concept of principal modes. We propose ``super-PMs'', which outperform PMs in terms of bandwidths, in section III. In section IV, we show that by reversing the concept of ``super-PMs'' , we achieve the ``anti-PM'', a state with much narrower bandwidths. 

\section{II. Principal Modes}

\begin{figure}[h]
	\begin{center}
		\includegraphics[angle=0, scale=1, width=0.8\columnwidth]{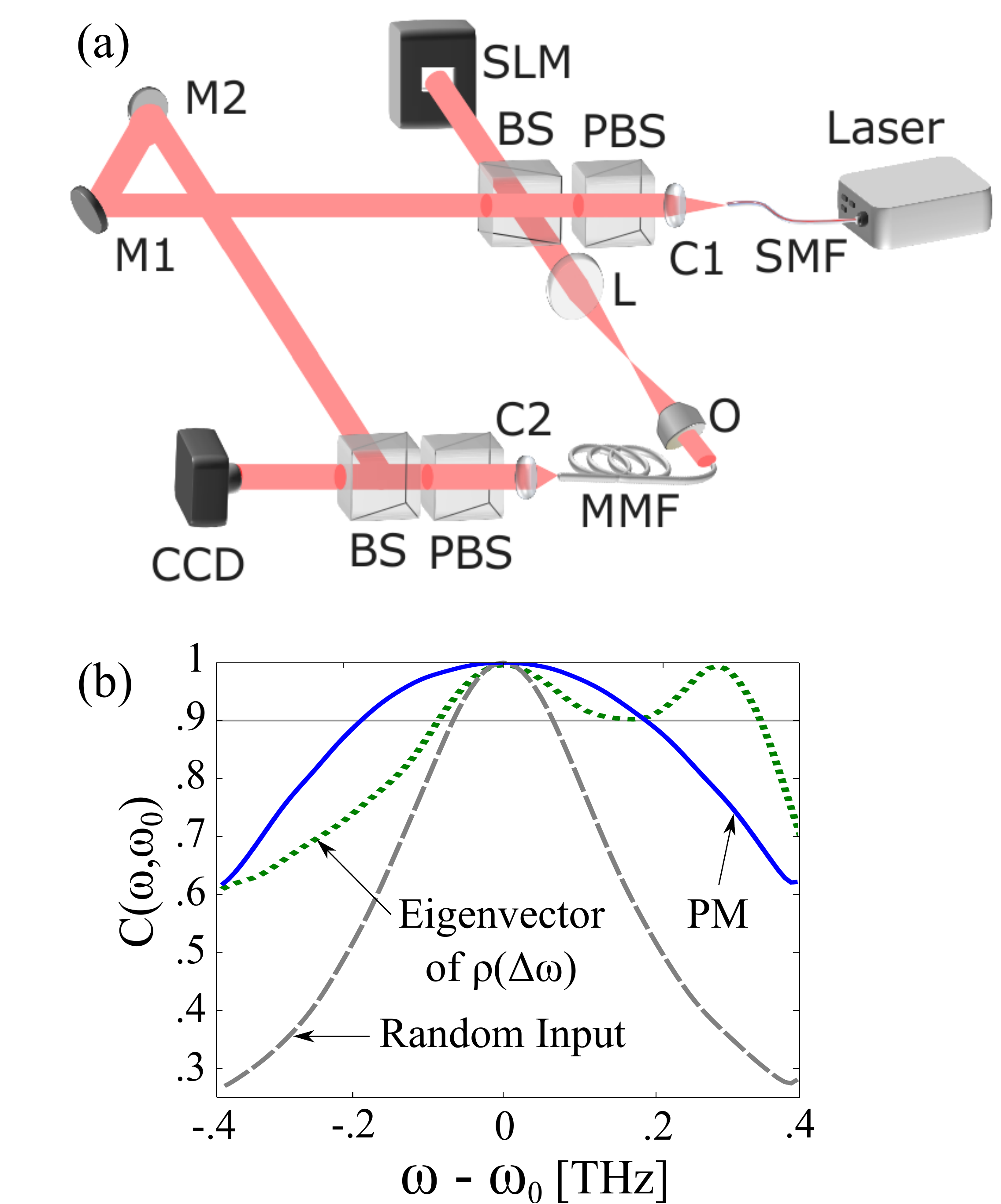}
		\caption{(a) Schematics of the experimental setup. The output from a frequency-tunable laser, injected by a single mode fiber (SMF), is collimated by a lens (C1) and linearly polarized by a polarization beam splitter (PBS) before being split by a beam splitter (BS) into two paths. (i) The signal is spatially modulated by the SLM, and then imaged onto the fiber facet of the MMF by a lens (L) and a micro objective (O). (ii) The reference arm matches the signal arm's pathlength by a pair of mirrors M$_1$ and M$_2$. The output from the MMF is collimated (C2) and polarized (PBS) before combining with the reference signal at BS with a relative angle. The interference fringes are recored by a CCD camera. (b) Measured autocorrelation function $\coreq$ for the output signals of the most stable PM (blue solid line), the average correlation for 20 random inputs (gray dashed line) and a super-PM eigenvector of $\rho \! \left(\Delta\omega \right)$ with $\Delta\omega = 0.28$ THz (green dotted line). The horizontal solid gray line indicates the threshold value $\coreq = 0.90$ that sets the correlation width. }
		\label{fig:PM}
	\end{center}
\end{figure}

Principal modes $\phat$ are the eigenstates of the Wigner-Smith time-delay operator $q$ for a MMF with transmission matrix $t(\omega)$, 
\begin{equation} \label{time-delay-eigenstate}
q \, \phat = -i \, t^{-1} \! \left( \omega_0 \right) \frac{dt}{d\omega}|_{\omega = \omega_0} \, \phat = \tau \, \phat,
\end{equation}
where $\tau$ is the respective complex eigenvalue whose real part is the so-called proper delay time \cite{Fan2005,Carpenter2015,Xiong2015, Ambichl2015}. From the above construction it directly follows that PMs do not suffer from modal dispersion to the first order in frequency change (see supplementary material \cite{Ambichl2015}), such that the output field pattern does not vary with frequency in the vicinity of $\omega_0$. A suitable measure for the modal dispersion, i.e., the deviation of the output vector $\psreq = \treq \phat$ from its respective direction at $\omega_0$, is the field autocorrelation function $\coreq := | \hat{\psi}^{\dagger} \! ( \omega ) \cdot \hat{\psi} ( \omega_0 ) | \in \left[ 0,1 \right]$, where $\psheq = \psreq \! / |\psreq| $ is the normalized output vector. We define the correlation width of a given input state as the spectral range surrounding $\omega_0$ where $C(\omega, \omega_0) \geq 0.9$ \cite{Xiong2015}.
 
Experimentally we tested a one-meter-long step-index MMF with a core diameter of 50 $\mu$m and a numerical aperture of 0.22. It features about 120 linearly polarized (LP) modes at $\lambda_0=1550$ nm for one polarization. To directly access the regime of strong mode coupling, we apply stress to the fiber through external clamps. A tunable laser is employed for measuring the transmission matrix at different frequencies.  Figure \ref{fig:PM}(a) shows our off-axis holographic setup for measuring the light field transmitted through the MMF. We use a spatial light modulator (SLM), which is imaged to the input facet of the MMF, to scan the incident angle of monochromatic laser light.  For each input wavefront, the output signal from the fiber combines with the reference arm, forming interference fringes on the camera. We extract the amplitude and phase of the field from the interference pattern. 

After recording the multi-spectral transmission matrix $\treq$, we truncated the matrix to reduce noise by discarding the eigenchannels with very high loss (90 independent transmission channels were retained out of the 120 available ones). Based on the truncated transmission matrix, we compute the time-delay matrix $q$ and find its eigenstates (PMs). Applying a computer-generated phase hologram to the SLM, we create the PMs by constructing their input states with amplitude and phase modulations \cite{arrizon2007pixelated,Bagnoud2004}. 
The output field patterns are recorded as the input frequency is scanned. Figure~\ref{fig:PM}(b) shows the measured auto-correlation function $C(\omega, \omega_0)$ for the PM with the largest correlation width (blue solid line), which has about three times the width as compared to that of a random input wavefront (gray dashed line). 

\section{III. Super-Principal Modes}

The fundamental question we address here is whether it is possible to increase the correlation width beyond the values obtained for PMs. To achieve this goal, we revisit the design principle of PMs, which are generated by the operator $q$ in Eq.~(\ref{time-delay-eigenstate}) such that their output field patterns do not change when changing the input frequency incrementally by $d\omega$ from the reference value $\omega_0$ (see supplementary material \cite{Ambichl2015}). In a first step we modify this approach to produce a state whose output pattern stays invariant when the input frequency is changed by a finite shift $\Delta \omega$ rather than by the infinitesimal value $d\omega$. 

To implement such a strategy, we introduce a new operator that aligns the output vectors of its eigenstate at two arbitrarily spaced frequencies $\omega_0$ and $\omega$ (with $\omega-\omega_0=\Delta\omega$). This new operator $\rhoreq$ takes the following form, 
\begin{equation} \label{rho-operator}
\rhoreq: = -i  \, t^{-1} \! \left( \omega_0 \right) \frac{t \! \left( \omega \right) - t \! \left( \omega_0 \right)}{\omega - \omega_0}\,.
\end{equation}
It approaches the time-delay operator $q$ for small frequency spacing between $\omega_0$ and $\omega$, i.e., $q=\lim_{\Delta\omega \rightarrow 0} \rho (\omega, \omega_0)$. An eigenstate of this new operator $\rhoreq$ features an output correlation function $\coreq$, which peaks not only at $\omega_0$ but also at $\omega$. With a gradual increase of $\left| \Delta\omega \right|$, the two correlation peaks move further apart, suggesting the possibility of extending the correlation bandwidth beyond that of the PM. We thereby use the term  ``super-PM'' to label a state that has a correlation bandwidth of the output field pattern exceeding that of the {\it widest} PM in a given fiber.

From the experimentally measured transmission matrix $\treq$,  we numerically generate the super-PMs as the eigenstates of $\rhoreq$ in Eq.~(\ref{rho-operator}), and find the anticipated two peaks at $\omega$ and $\omega_0$ in the correlation function, as shown by the green dotted curve in Fig.~\ref{fig:PM}(b). The width of each peak is given approximately by the spectral correlation width $\delta \omega$ of the transmission matrix (i.e., the correlation width associated with random input). 
 As the frequency spacing between the two peaks at $\omega$ and $\omega_0$ exceeds the peak width $\delta \omega$, a dip develops in between the two peaks. To maximize the correlation bandwidth, we choose the spacing $\omega -\omega_0$ such that $C(\omega, \omega_0)$ drops to the threshold value of 0.9 in between the two peaks in Fig.~\ref{fig:PM}(b). In this case, the correlation width is about $18 \%$ wider than the width of the widest PM. A further increase of $\Delta\omega$ makes the dip in between the two peaks drop to below $0.9$, causing a sudden decrease of bandwidth. 

To reach the ideal case of a flat correlation function $\coreq\approx 1$ over the entire interval $[\omega_0,\omega]$, $\Delta\omega$ must be less than $\delta \omega$. The fundamental reason that prevents such a flat correlation curve from being realized in a broader frequency interval can be understood from the construction principle of $\rhoreq$ and of its eigenstates: a state with a flat correlation function $\coreq \equiv 1$ in a finite interval $[\omega_0,\omega]$ would have to be a simultaneous eigenstate of \textit{all} operators $\rhoreq$ in this $\omega$-interval. The relevant commutator that would have to vanish for this to be possible is, however, non-zero in general: $[\rho (\omega_1, \omega_0), \rho (\omega_2, \omega_0)]\neq 0$ with $|\omega_1 - \omega_2| > \delta \omega$. 

\begin{figure*}[!tb]
  \begin{center}
    \includegraphics[angle=0, scale=1.0, width=\textwidth]{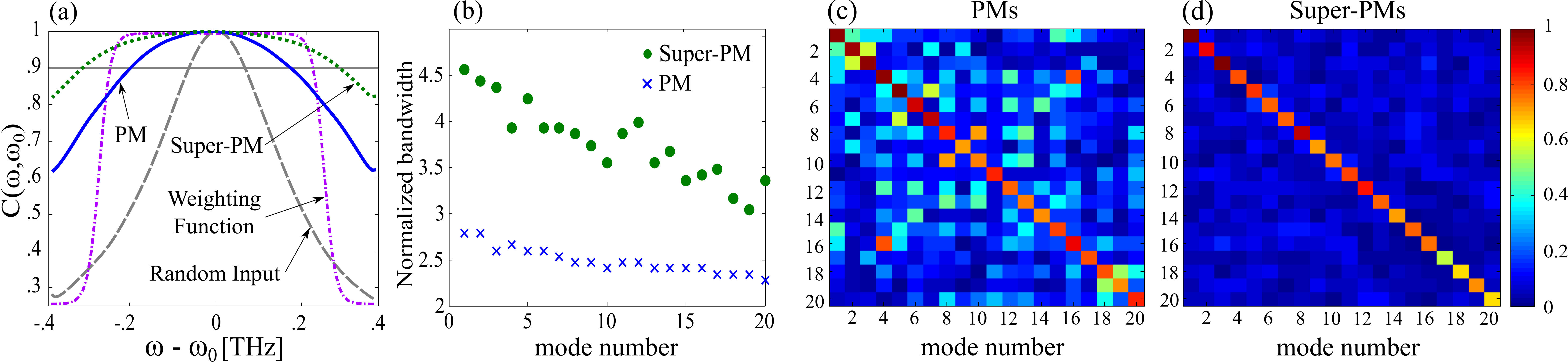}
    \caption{\textcolor{black}{ (a) Measured spectral correlation function $\coreq$ for the output signals of the widest PM (blue solid line), and for the widest super-PM obtained from our bandwidth optimization (green dotted line).  The weighting function $\wheq$ used in the cost function of Eq.~(\ref{trans-cost}) is shown as purple dash-dotted line. The gray dashed line shows the spectral correlation of a random input. The horizontal solid gray line indicates the threshold value $\coreq = 0.90$ that sets the correlation width.  (b) Spectral correlation width of super-PMs and PMs normalized by that of a random input. Our optimization scheme found 20 super-PMs that have a larger bandwidth than the widest PM.
The orthogonality of the output field patterns for the 20 widest PMs (c) and the 20 super-PMs (d). In the presence of loss, the PMs are not orthogonal but the super-PMs are. 
    } }
    \label{fig:OSPMt}
  \end{center}
\end{figure*}

In spite of this restriction, we will now show how to create a set of states with correlation bandwidth considerably exceeding that of conventional PMs. In addition, we will make such  ``super-PMs'' mutually orthogonal, even when the fiber has MDL. Both the PMs and super-PMs described above are non-orthgonal, as the operators $q(\omega_0)$ and $\rhoreq$ become non-Hermitian in the presence of MDL.
We will thus work with a new approach that constructs the desired set of super-PMs as {\it approximate} rather than as {\it perfect} simultaneous eigenvectors of the operators $\rhoreq$ within a desired spectral interval. Achieving this objective calls for the implementation of an optimization procedure with a non-linear cost function involving the multi-spectral transmission matrix $\treq$ and the operators $\rhoreq$. Given the dimensionality of the problem determined by the number of fiber modes and the width of the spectral region of interest, such an optimization is a very demanding task even numerically. 

In the following we take a shortcut to create orthogonal super-PMs by working with a considerably reduced optimization functional $\mathcal{T} $ (cost function),
\begin{eqnarray} \label{trans-cost}
\mathcal{T} \! \left( \phat \right) & := & \int\! d\omega\,\left[1 - C \! \left( \omega, \omega_0 \right)^2 \right] \wheq \nonumber\\
 & = & \int\! d\omega \,\left[1 - \frac{\left|\vec{\psi}^{\dagger}\!\left(\omega\right)\cdot\vec{\psi}\!\left(\omega_{0}\right)\right|^2} {\left|\vec{\psi}\!\left(\omega\right)\right|^2\left|\vec{\psi}\!\left(\omega_{0}\right)\right|^2} \right] \wheq.
\end{eqnarray}

The spectral range that is effectively taken into account in this functional is defined by the weighting function $\wheq$, which is chosen to be a step-like, but continuous function centered around $\omega_0$, see Fig.~\ref{fig:OSPMt}(a) (purple dash-dotted curve). To maximize the width of the correlation function $\coreq$, the (scalar) value of $\mathcal{T}$ needs to be minimized depending on the input state $\phat = t^{-1}\!\left(\omega\right) \psreq$. We implement an efficient minimization of $\mathcal{T}$ using a simple gradient-based scheme (see supplemental material \cite{Ambichl2015} for the exact expressions of $W(\omega)$ and $\delta \mathcal{T} / \delta \phat$). We note that our optimization scheme is a purely numerical optimization based on the experimentally measured transmission matrices rather than an experimental feedback loop.

We typically obtain the widest super-PM with an optimization loop that is initiated with the widest PM as a starting point for the corresponding iteration. Next, a whole cascade of optimizations of Eq.~(\ref{trans-cost}) is carried out, where in each optimization loop for a single super-PM the cost function $\mathcal{T} $ is minimized in the vector space orthogonal to each of the super-PMs obtained in the preceding steps. In this way, a strictly orthogonal set of super-PMs is obtained. 

The input states for these super-PMs are created by the SLM and injected into the fiber. Figure~\ref{fig:OSPMt}(a) shows the widest super-PM (green dotted line). Its spectral correlation width is about $70\%$ larger than that of the widest PM.  Out of the total 90 modes in the fiber after the matrix truncation, we find 20 super-PMs that outperform the widest PM in terms of the correlation width.  Figure~\ref{fig:OSPMt}(b) shows the spectral width of super-PMs normalized by that of random input wavefronts. The widest super-PM has a bandwidth increase of 4.5 times as compared to the random input. In addition, we also plot the 20 PMs with largest bandwidth and find that the bandwidth enhancement for super-PMs is well above that of the PMs.

Due to fiber absorption and scattering loss, the PMs are non-orthogonal already at the fiber input, since they are the eigenstates of the non-Hermitian operator $q$. The propagation through the fiber will further increase this non-orthogonality, thus PMs are non-orthogonal both at the fiber input and output ends. By contrast, the super-PMs, in the way we construct them, are orthogonal at the input. We also observe that the orthogonality of the super-PMs at the input facet is well sustained during propagation through the fiber. The overlap of the output field patterns is defined as $O_{mn} := | \hat{\psi}^\dagger_m \left( \omega_0 \right) \cdot \hat{\psi}_n \left( \omega_0 \right) |$ for the $m$-th and $n$-th super-PM. For a perfectly orthonormal set of vectors we would have $O_{mn} = \delta_{mn}$, where $\delta_{mn}$ is the Kronecker delta. The 20 widest PMs we consider are rather far away from this ideal case, as shown in Fig.~\ref{fig:OSPMt}(c). The average off-diagonal element  of $O$ is 0.188. The super-PMs preserve their initial orthogonality to a large extent as shown in Fig.~\ref{fig:OSPMt}(d). On average we find for the off-diagonal elements of $O$ a reduced value of $0.068$. Thus the crosstalk between different super-PMs is strongly suppressed as compared to the PMs.  

To validate our initial hypothesis that the enhanced bandwidth of super-PMs is linked to the property that they mimic mutual eigenstates of all $\rhoreq$ operators in the spectral region of interest, we work out a corresponding measure that quantifies how close an input state is to a simultaneous eigenvector of these operators (see supplemental material \cite{Ambichl2015} for details). Applying this measure to the PMs and to the super-PMs, indeed, shows that the super-PMs are much closer to mutual eigenstates of $\rhoreq$ than the conventional PMs, despite the fact that the information on $\rhoreq$ is included in the cost function $\mathcal{T}$ only implicitly (see supplemental material \cite{Ambichl2015} for the corresponding numerical values we obtained).

\begin{figure}[h]
  \begin{center}
    \includegraphics[angle=0, scale=1.0, width=\columnwidth]{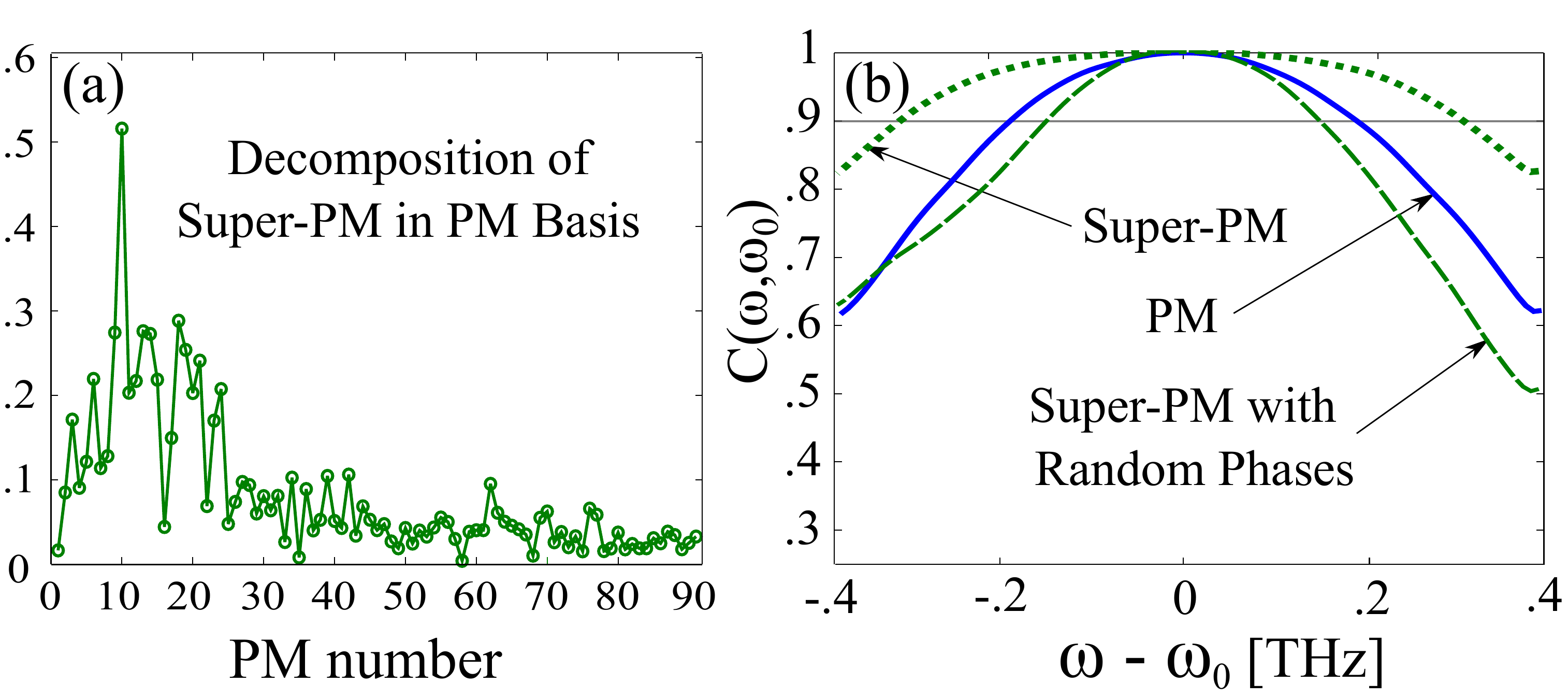}
    \caption{\textcolor{black}{ (a) Decomposition of the widest super-PM in the basis of PMs, which are numbered from short to long delay time. The super-PM bundles together PMs with similar delay times. (b) Spectral correlation functions of the widest super-PM (green dotted line), the widest PM (blue solid line) and the widest super-PM with phases of constituent PMs randomized (green dashed line). The bandwidth enhancement of super-PMs is sensitive to the phases of constituent PMs, indicating that super-PMs are formed by interference of PMs.} }
    \label{fig:fig3}
  \end{center}
\end{figure}

Can we also understand on a more intuitive level why and how super-PMs manage to outperform PMs? To answer this question we first decompose both the PMs as well as the super-PMs realized in the experiment in the basis of LP modes of the fiber. In the regime of strong mode coupling where all of the results shown above were obtained, we find that both PMs and super-PMs consist of nearly all LP modes, and the higher-order LP modes have smaller contributions due to higher loss (see supplemental material \cite{Ambichl2015}). While this finding confirms that PMs as well as super-PMs are really non-trivial combinations of fiber modes in the strong-coupling regime \cite{Xiong2015}, it does not shed any light on the difference between PMs and super-PMs. We therefore change basis from the LP-modes to the PMs themselves, which constitute the natural basis to capture the dynamical aspects of light scattering as each PM is associated with a proper delay time. Also their greatly reduced wavelength dependence makes PMs very suitable to describe the transmitted light in a broad frequency window. When decomposing the  super-PMs in the basis of PMs (with the bi-orthogonal basis vectors being sorted according to the delay times of the PMs), we observe clearly that the super-PMs are composed of PMs with neighboring delay times, as shown in Fig.~\ref{fig:fig3}(a). 

The physical interpretation of this result is that super-PMs build on a very narrow distribution of delay times and enhance the spectral correlation width via interference of several time-delay eigenstates. To demonstrate that the relative phases of constituent PMs are essential, we also plot in Fig.~\ref{fig:fig3}(b) the correlation functions after adding random phases to the decomposition coefficients of the PMs (while maintaining the absolute magnitudes). It is evident that randomizing the phase significantly decreases the correlation width. The sensitivity to the relative phase of constituent PMs indicates that the interference of the PMs involved in the formation of a super-PM is essential.

Another relevant aspect is that the mixing of neighboring PMs to create super-PMs requires that a sufficient number of PMs are available with similar delay times. In a MMF with weak mode coupling, the values of the delay times are spread over a much broader time range, and hence the super-PM is composed of only one or two PMs with neighboring delay times. Therefore, the bandwidth of a super-PM is typically not significantly different from its constituent PMs for weak mode coupling. To test this conjecture, we apply the same optimization algorithm as above to fibers in the weak mode coupling regime. The corresponding results indicate that in this case the super-PMs are very similar to individual PMs and no significant admixture from several PMs is seen. Consequently, the correlation width of even the widest super-PM barely exceeds that of the widest PM (see the supplemental material \cite{Ambichl2015}). Super-PMs thus realize their full potential in the regime of strong-mode coupling where a gain in bandwidth is also most relevant since the PM-bandwidths in the strong-coupling limit are much narrower than for weak mode-coupling \cite{XiongOE17}.

\section{IV. Anti-Principal Modes}

We will now investigate whether novel states of light may also be found when turning the concept of principal modes on its head. Specifically, we will address the question, whether it is possible to create not only states with a very broad spectral correlation width such as the PMs or the super-PMs studied above, but also states with a drastically reduced bandwidth as compared to the values associated with a typical or random input wavefront. Such ``anti-PMs'' would have an enhanced frequency sensitivity as desired, e.g., for fiber-based spectrometers whose operation principle relies on a very narrow correlation bandwidth \cite{redding_all-fiber_2013,redding_using_2012}. The optimization algorithm presented above now allows us to generate such highly sensitive states by just maximizing instead of minimizing the functional $\mathcal{T} $ in Eq.~(\ref{trans-cost}).

\begin{figure}[t]
  \begin{center}
    \includegraphics[angle=0, scale=1.0, width=\columnwidth]{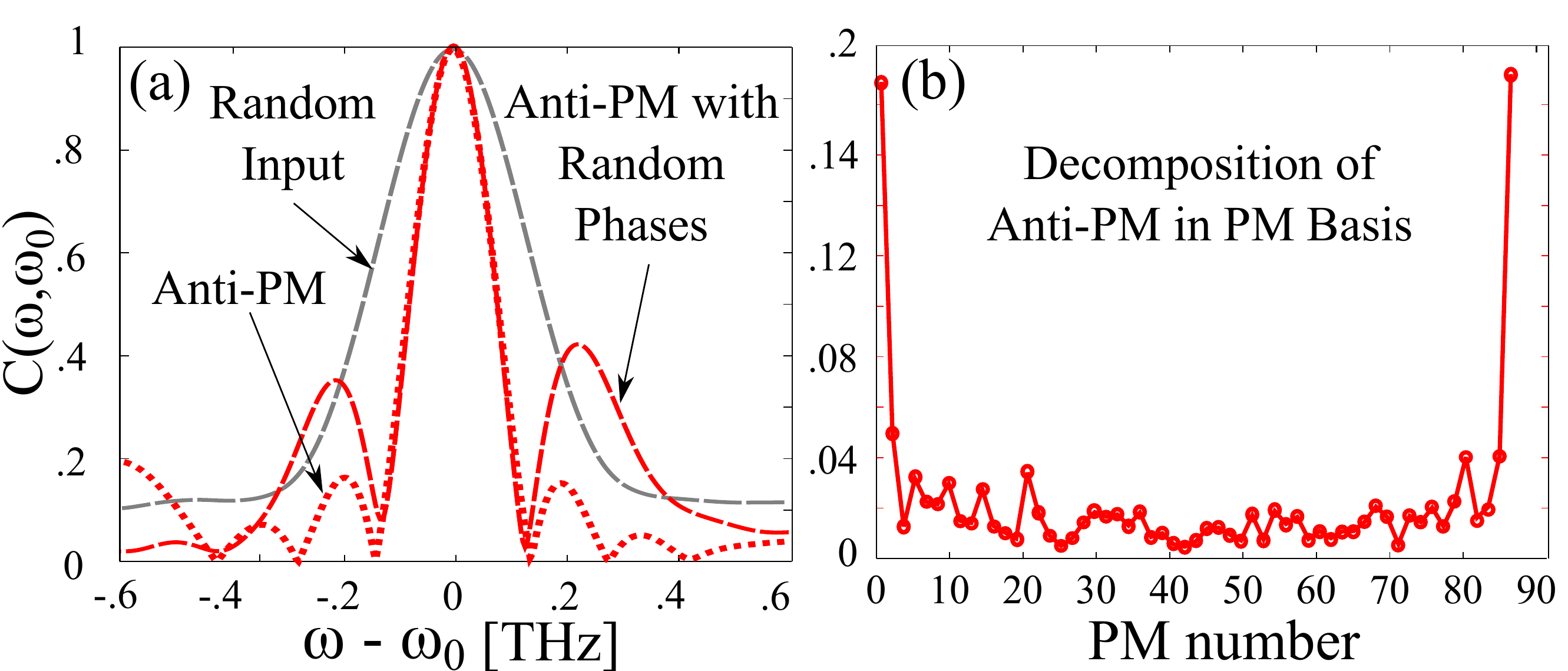}
    \caption{\textcolor{black}{Numerically determined anti-PM in a fiber without MDL. (a) Spectral correlation function of the narrowest anti-PM (red dotted line), which is significantly narrower than that of the random input (gray dashed line). The red dashed line is obtained by randomizing the phases of constituent PMs in the anti-PM. The phase randomization does not change the main peak, but enhances the two side peaks. (b) The decomposition of this anti-PM in the PM basis shows two pronounced maxima at the extreme values of the involved PMs, indicating that anti-PMs are efficiently formed by combining the fastest and the slowest PMs.  The data shown here result from the numerical simulation of a waveguide with 200 $\mu$m width, 0.22 numerical aperture, 1 meter length and no loss.} }
    \label{fig:fig4}
  \end{center}
\end{figure}

We first apply the algorithm for anti-PMs on an ideal MMF with no MDL. The corresponding unitary transmission matrix is obtained numerically with the concatenated fiber model developed earlier \cite{hokahn2011}. For simplicity, we consider a planar waveguide with a core width of 200 $\mu m$ and a numerical aperture of 0.22, supporting 57 guided modes \cite{Okamoto}. The total length of the waveguide is one meter, divided into 20 segments in each of which light propagates without mode coupling. Between adjacent segments, all modes are randomly coupled, as simulated by a unitary random matrix \cite{XiongOE17}. The results obtained by applying the optimization algorithm to this numerical model are shown in Fig.~\ref{fig:fig4}(a), displaying a significant reduction of the spectral correlation width of the anti-PM as compared to random inputs. Moreover, when decomposing the anti-PM in the basis of PMs, we immediately see that the narrowest anti-PM consists mainly of admixtures between the fastest and slowest PMs [see Fig.~\ref{fig:fig4}(b)]. These PMs have the largest achievable temporal difference in the MMF and mixing them with similar weight thus leads to an efficient decorrelation when changing the input frequency. Quite intuitively, the anti-PMs thus not only form the antipodes of the super-PMs in terms of their correlation bandwidth, but also in terms of the way they are constructed: while super-PMs group together several PMs with similar time delay, the anti-PMs combine those PMs with the most different time delay available. Whereas the relative phase with which this superposition of PMs is implemented clearly matters for super-PMs [see Fig.~\ref{fig:fig3}(b)], we find that for anti-PMs the correlation bandwidth is barely affected when changing the phases of constituent PMs [see the red dashed curve in Fig.~\ref{fig:fig4}(a)]. 

The reason behind this observation is that for PMs associated with very different delay times adjusting their relative phase cannot mend these states' intrinsic tendency to decorrelate with frequency, while for states with nearby delay times the appropriate phases will optimize the interferences for specific mode-coupling in the MMF such as to increase the correlation width. As shown by the red dashed curve in Fig.~\ref{fig:fig4}(a), the correlation function features two enhanced side peaks adjacent to the main peak for the anti-PM with phases of constituent PMs randomized. After reaching zero, the correlation revives as $\omega$ is tuned further from $\omega_{0}$. This revival is caused by the beating of the slowest and fastest PMs, the two main components of the anti-PM. When the relative phase of the fastest and slowest PMs accumulates 2$\pi$, the output field should be the same as the one when the relative phase is zero. However, small contributions from PMs with intermediate delay times give rise to multi-path interference that suppresses the side peaks in the anti-PM in a way that is phase-sensitive.

To connect with the experimental case of anti-PMs, it is essential to take into account the mode dependent loss (MDL) in the fiber. 
%We know already that MDL renders conventional PMs non-orthogonal to each other - a feature that we remedied in the design of super- and anti-PMs by choosing them to be orthogonal to each other at the input in the optimization procedure. MDL, however, also has other interesting consequences on our optimized states \comment{please discuss them here.}
We gradually increase the MDL in the numerical simulations up to the value close to the MMF used in the experiment. The narrowest anti-PM displays a narrowing of the spectral correlation function, and its bi-modal composition in the PM-basis changes drastically (see supplemental material \cite{Ambichl2015}).  In particular, the increase of MDL enhances the weight of the slowest PMs, and the composition of the anti-PM in the PM-basis becomes much broader. This can be explained by considering that the MDL attenuates the slowest PMs most strongly since they stay longest inside the fiber. The broadening of the PM-composition can then be understood as a mechanism to compensate for the reduced contributions from the slow PMs. Since the constituent PMs with similar delay times overlap in time, their relative phases become important. When randomizing their phases, the bandwidth of anti-PMs is increased (see supplemental material \cite{Ambichl2015}).

\begin{figure}[t]
  \begin{center}
    \includegraphics[angle=0, scale=1.0, width=\columnwidth]{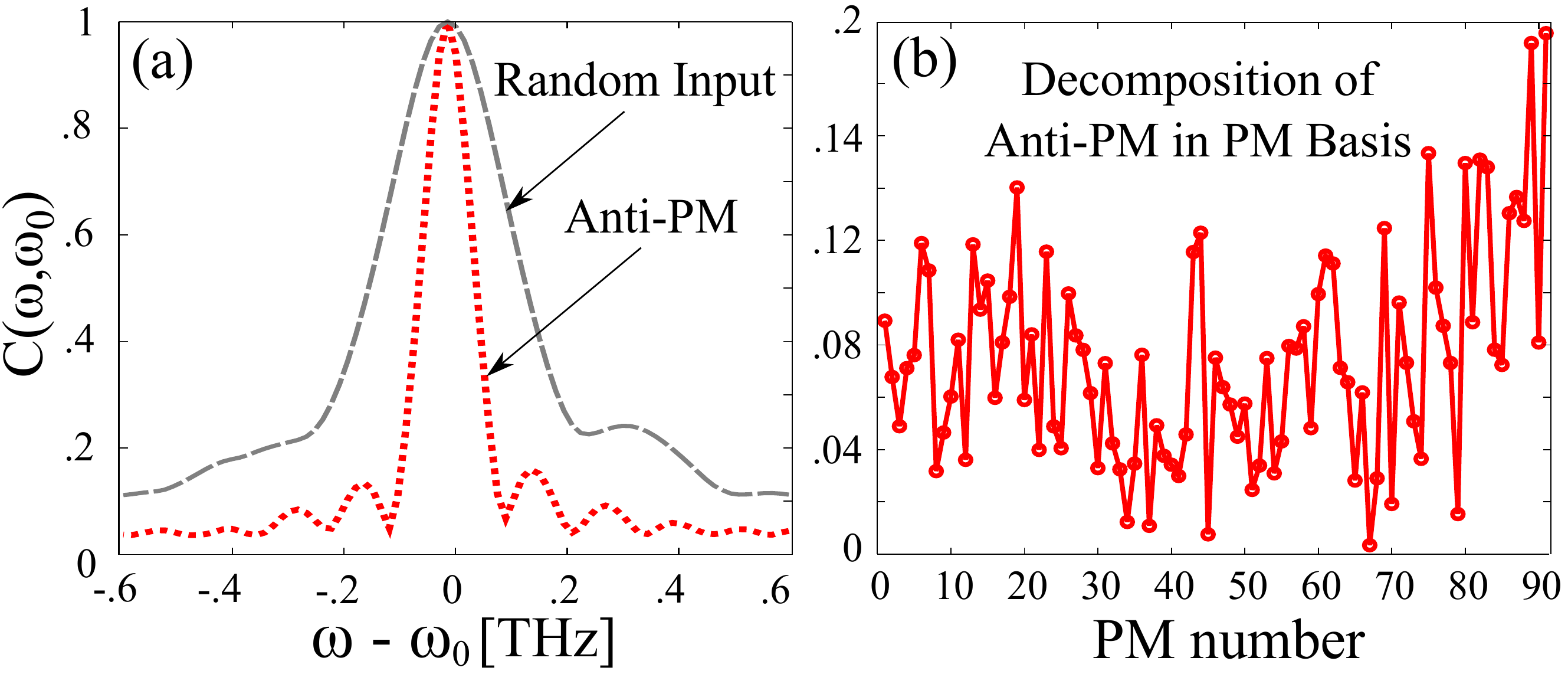}
    \caption{\textcolor{black}{ Experimentally measured anti-PM in the same fiber as in Fig. 1. (a) The spectral correlation function of the narrowest anti-PM (red dotted line), which is notably narrower than that for a random input (gray dashed line). (b) Due to mode-dependent loss, the decomposition of this anti-PM in the PM basis does not give the bi-modal distribution as in Fig.~\ref{fig:fig4}(b), but rather a broad distribution over all available PMs. This result agrees with the numerical simulation that includes the MDL (see the supplemental material \cite{Ambichl2015}). } }
    \label{fig:fig5}
  \end{center}
\end{figure}

When generating the anti-PMs in the MMF we find  62 anti-PMs out of 90 fiber modes, whose spectral correlation width is narrower than that of random inputs. Fig.~\ref{fig:fig5} shows the data for the narrowest anti-PM. Due to the MDL, it is broadly spread in the PM basis,  but it still decorrelates much more quickly than the random input with frequency detuning. Only when moving from the strong to the weak mode-coupling limit do we find that anti-PMs have about the same correlation width as a random input even without MDL (see supplemental material \cite{Ambichl2015}). This is because in the weak mode coupling regime, the fastest PM and the slowest PM have very different delay times. The beating between the two PMs is so strong that it causes significant side peaks of the spectral correlation function. To reduce the side peaks, additional PMs with intermediate delay times must be included. The inclusion of many intermediate PMs makes the correlation width of the anti-PM similar to that of a random input. Therefore, both super-PMs and anti-PMs have in common that they unfold their full potential in the limit of strong mode coupling. 

\section{V. Summary}

In summary, we present a new set of light states in multi-mode fibers that have either a significantly broader spectral correlation than the principal modes or a significantly narrower spectral correlation than a random input wavefront, respectively. We thus term these new states super- and anti-principal modes, and demonstrate how to generate them with a simple gradient-based algorithm based on the experimentally measured multi-spectral transmission matrix. Our optimization algorithm allows us to determine a whole set of such super- or anti-PMs, which are mutually orthogonal to each other even in the presence of mode-dependent loss and thus feature much reduced crosstalk. By overcoming the limitations of PMs, super-PMs outperform the PMs in terms of bandwidth and orthogonality. The performance gain is the highest in the regime of strong mode coupling as for very long fibers, where an increase in bandwidth is also most sought-after. 

Moreover, we provide a physical understanding of how these states are formed by decomposing them in the PM basis. Super-PMs tend to combine several PMs with nearby delay times in a super-position with optimized phases. Anti-PMs, on the contrary, tend to combine PMs with the most different delay times available in the fiber. The presence of MDLs leads to modifications of these states, which are analyzed in detail. 

On the one hand, the large bandwidth and mutual orthogonality of super-PMs pave the way for their application to dispersion-free transmission of light pulses through multi-mode fibers. On the other hand, the high spectral sensitivity of anti-PMs is ideally suited for optimizing the resolution of fiber-based spectrometers as well as for reducing the spatial coherence of a broadband light source for crosstalk-free imaging. Furthermore, this work illustrates the potential of using PMs to synthesize novel light states beyond the ones demonstrated here.

\section{Acknowledgment}

We acknowledge Chia Wei Hsu, Joel Carpenter, and Nicolas Fontaine for helpful discussions. This work is supported partly by the U.S. National Science Foundation under the Grant No.ECCS-1509361 and by the US Office of Naval Research under the MURI grant No. N00014-13-1-0649.
P. A. and S. R. acknowledge support by the Austrian Science Fund (FWF) through projects SFB NextLite (F49-P10) and by the European Commission under project NHQWAVE (GA number 691209).

\bibliographystyle{apsrev4-1}
\bibliography{OSPM}
\end{document}

\setlength{\tabcolsep}{1pt}
\title{Super- and Anti-Principal Modes in Multi-Mode Waveguides}

\author{Philipp Ambichl}
\affiliation{Institute for Theoretical Physics,
Vienna University of Technology, A-1040, Vienna, Austria, EU}
\author{Wen Xiong}
\affiliation{Department of Applied Physics,
Yale University, New Haven, Connecticut 06520, USA}
\author{Yaron Bromberg}
\affiliation{Department of Applied Physics,
Yale University, New Haven, Connecticut 06520, USA}
\author{Brandon Redding}
\affiliation{Department of Applied Physics,
Yale University, New Haven, Connecticut 06520, USA}
\author{Hui Cao}
\affiliation{Department of Applied Physics,
Yale University, New Haven, Connecticut 06520, USA}
\author{Stefan Rotter}
\affiliation{Institute for Theoretical Physics,
Vienna University of Technology, A-1040, Vienna, Austria, EU}

\maketitle
\section{A: Projected transmission matrix and its effective inverse}
For the construction of the time delay operator $q = -i \, t^{-1} \!\left( \omega_0 \right) dt / d\omega|_{\omega_0}$, the inverse of the transmission matrix $\treq$ at $\omega_0$ is needed. A straight-forward inversion is generally not possible, because the matrix $t \!\left( \omega_0 \right)$ may be singular or not quadratic. In the following, we present an algorithm that produces an effective inverse, irrespective of the shape or singularity of $t \!\left( \omega_0 \right)$, that preserves the desired features like the broad correlation widths of the principal modes. In a first step we perform a singular value decomposition (SVD) of the transmission matrix at the reference frequency,
\begin{equation}\label{SVD}
t \!\left( \omega_0 \right) = U \, \Sigma \, V^\dagger,
\end{equation}
where the matrices $U$ and $V$ contain column-wise the left and right singular vectors, respectively. The diagonal matrix $\Sigma = {\rm diag} \! \left( \left\{ \sigma_n \right\} \right)$ contains the (real and positive) singular values $\sigma_n$ of $t \!\left( \omega_0 \right)$. In order to eliminate possible singularities, or more generally, ``dark'' channels that contribute only weakly to the total output speckle, we only keep those $N_{\rm bright}$ singular values $\left\{ \sigma_n > \epsilon \right\}$, and the corresponding left and right singular vectors. In order to achieve the results presented in the main text, we choose the value of the threshold $\epsilon$ to be equal to 10\% of the maximum singular value. This selection leads to matrices of reduced dimensions, $u$, $v$, and $\sigma = {\rm diag} \! \left( \left\{ \sigma_n > \epsilon \right\} \right)$. The next step is to project the transmission matrix onto the resulting reduced subspace of ``bright'' channels,
\begin{equation}
t_{\rm bright} \! \left( \omega_0 \right) = u^\dagger \, t \!\left( \omega_0 \right) \, v,
\end{equation}
which yields the non-singular, quadratic and, therefore, invertible $N_{\rm bright} \times N_{\rm bright}$ matrix $t_{\rm bright} \!\left( \omega_0 \right)$. The effective inverse is then finally obtained by inversion and reprojection onto the original vector space,
\begin{eqnarray}\label{effective-inverse}
t_p^{-1} \!\left( \omega_0 \right) & := & v \, t^{-1}_{\rm bright} \!\left( \omega_0 \right) \, u^\dagger \nonumber \\
& = & v \, \left( u^\dagger \, t \!\left( \omega_0 \right) \, v \right)^{-1} \, u^\dagger.
\end{eqnarray}
We found that a projection onto the subspace of bright channels is generally beneficial in terms of eliminating the noisy background of the output signal, not only in the construction of the time delay operator, but also for the cost functions in our optimization scheme. For that reason, we also introduce the projected transmission matrix,
\begin{equation}\label{t-projected}
t_{p} \! \left( \omega \right) = u \, u^\dagger \, t \! \left( \omega \right) \, v \, v^\dagger,
\end{equation}
that we use for the construction of the operator $q \! \left( \omega_0 \right)$, $\rhoeq$, and the cost functions $\mathcal{T} ( \phat )$ as well as $\mathcal{L} ( \phat )$, discussed in the main text. Since $u^\dagger \, u = 1$ and $v^\dagger \, v = 1$, the effective inverse we defined in Eq.~(\ref{effective-inverse}) and the projected matrix in Eq.~(\ref{t-projected}) satisfy $t^{-1}_{p} \! \left( \omega_0 \right) t_{p} \! \left( \omega_0 \right) t^{-1}_{p} \! \left( \omega_0 \right) = t_{p} \! \left( \omega_0 \right)$, and $t^{-1}_{p} \! \left( \omega_0 \right) t_{p} \! \left( \omega_0 \right) t^{-1}_{p} \! \left( \omega_0 \right) = t^{-1}_{p} \! \left( \omega_0 \right)$ for pseudo-inverse matrices. The products of the two matrices give $t_{p} \! \left( \omega_0 \right) t^{-1}_{p} \! \left( \omega_0 \right) = u \, u^\dagger$ and $t^{-1}_{p} \! \left( \omega_0 \right) t_{p} \! \left( \omega_0 \right) = v \, v^\dagger $, respectively, where the projectors $u \, u^\dagger$ and $v \, v^\dagger$ can be viewed as the representations of the identity matrices in the bright left and right subspaces the transmission matrix is projected onto.

\section{B: Vanishing derivative of output field pattern for the principal modes}
To demonstrate the property of a vanishing frequency derivative of the output speckle pattern if the input state is a PM (see also \cite{Fan2005}), we approximate the output vector with a Taylor series up to first order, $\psreq \approx \vec{\psi}\!\left(\omega_0\right) + d\vec{\psi} / d\omega |_{\omega = \omega_0} \left( \omega - \omega_0 \right)$. In a next step, we demand the first order term in this series to be aligned to the zero-th order term, which translates into the requirement that these two vectors are proportional to each other with a complex constant $\tau$,
\begin{equation} \label{PM_prop}
i \, \tau \, \vec{\psi}\!\left(\omega_0\right) =  \frac{d\vec{\psi}}{d\omega}|_{\omega = \omega_0}.
\end{equation}
Inserting the input-output relation $\psreq = \treq \phat $ into Eq.~(\ref{PM_prop}) and rearranging the terms, we get
\begin{eqnarray} \label{PM_qeigenstate}
\frac{dt}{d\omega}|_{\omega = \omega_0} \phat & = & i \, \tau \, t \!\left( \omega_0 \right) \, \phat \nonumber \\
-i \, t^{-1} \!\left( \omega_0 \right) \frac{dt}{d\omega}|_{\omega = \omega_0} \, \phat & = & \tau \, \phat,
\end{eqnarray}
directly showing that such a state is an eigenvector of the time delay operator $q = -i \, t^{-1} \!\left( \omega_0 \right) dt / d\omega|_{\omega_0}$, i.e. a PM. The constant $\tau$ plays the role of the corresponding eigenvalue.

\iffalse
\section{B Wavepacket envelope}
In order to produce a spectral envelope for the input pulse $\phreq$ that is continuous on the one hand, but similar to a rectangular function on the other hand, we utilize step-like Fermi functions of the following form,
\begin{equation} \label{Fermi-step}
f^\pm \! \left( \omega, \omega_0, b, s \right) = \frac{1}{1 + e^{-\left( \omega - \omega_0 \pm \,b/2 \right) / s}}.
\end{equation}
The actual envelope is then given by
\begin{equation}
\phi \! \left( \omega \right) = \frac{1}{N_\phi} \left( f^+ \left( \omega, \omega_0, b, s \right) + f^- \left( \omega, \omega_0, b, s \right) \right),
\end{equation}
where the parameter $\omega_0$ denotes the center frequency, $b$ is the FWHM of the envelope, and $s$ determines the steepness of the slopes. For $s=0$, the expressions in Eq.~(\ref{Fermi-step}) turn into discontinous step functions. The normalization constant is given by $N_\phi = \int \! d\omega \left| \phreq \right|^2$. For the experimental results displayed in the main text, we used the specific values of $b = 0.875$ THz and $s = 0.022$. For the weighting function $\wheq$, which is also discussed in the main text, we particularly used
\begin{equation}
\wheq = \left| \phreq \right|^2.
\end{equation}
\fi

\section{C: eigenstates, variance and expectation value of the $\rho$-operator}
The equation for an eigenstate of $\rhoeq$ reads
\begin{equation} \label{rho-eigenstates}
\rhoeq \phat = -i \, t^{-1} \! \left( \omega_0 \right) \frac{t \! \left( \omega \right) - t \! \left( \omega_0 \right)}{\omega - \omega_0} \phat = r \, \phat,
\end{equation}
with the complex eigenvalue $r$. In close analogy to the derivation in section B, we can rearrange Eq.~(\ref{rho-eigenstates}) like
\begin{eqnarray}
\frac{t \! \left( \omega \right) - t \! \left( \omega_0 \right)}{\omega - \omega_0} \phat & = & i \, r \, t \! \left( \omega_0 \right) \phat \nonumber \\
\frac{\psreq - \vec{\psi} \! \left( \omega_0 \right)}{\omega - \omega_0} & = & i \, r \, \vec{\psi} \! \left( \omega_0 \right) \nonumber \\
\Delta\psreq & = & i \, r \, \Delta\omega \, \vec{\psi} \! \left( \omega_0 \right) \nonumber \\
\Delta\psreq & = & \tilde{c} \, \vec{\psi} \! \left( \omega_0 \right). \label{recurrence}
\end{eqnarray}
Eq.~(\ref{recurrence}) states that for an eigenstate of $\rhoeq$, the deviation of the output vector at $\omega \ne \omega_0$, $\Delta\psreq = \psreq - \vec{\psi} \! \left( \omega_0 \right)$  is aligned to $\vec{\psi} \! \left( \omega_0 \right)$, so is $\psreq$. The  proportionality constant $\tilde{c} = i \, r \, \Delta\omega$.

For our purposes, we define the expectation value of $\rhoeq$ for a state $\phat$ analogously to that of a Hermitian operator like
\begin{equation}
\left\langle \rhoeq \right\rangle = \phat^\dagger \rhoeq \phat.
\end{equation} 
In contrast to a Hermitian operator, this expectation value is a complex number in general. The variance for the non-Hermitian operator $\rhoeq$, however, can be computed,
\begin{eqnarray}
\langle | \rhoeq - \left\langle \rhoeq \right\rangle |^2 \rangle & = & \phat^\dagger \rhoeq^\dagger \rhoeq \phat \nonumber \\
& - & | \phat^\dagger \rhoeq \phat |^2,
\end{eqnarray}
and gives a purely real number. Note that for an eigenstate of $\rhoeq$, the expectation value is equal to the corresponding (complex) eigenvalue, and the variance vanishes.

\section{D: the gradient-based optimization scheme}
The gradient with respect to $\phat^\dagger$ using $\psreq = \treq \phat$ for the cost function 
\begin{eqnarray}
\mathcal{T} \! \left( \phat \right) & = & \int\! d\omega \,\left(1 - \frac{\left|\vec{\psi}^{\dagger}\!\left(\omega\right) \cdot \vec{\psi} \! \left(\omega_{0}\right)\right|^2} {\left|\vec{\psi}\!\left(\omega\right)\right|^2 \left|\vec{\psi}\!\left(\omega_{0}\right)\right|^2} \right) \wheq \nonumber \\
& = & \int\! d\omega \,\left(1 - \frac{\left| \phat^\dagger t^\dagger \! \left(\omega\right) t \!\left(\omega_0\right) \phat \right|^2} {\left| \treq \phat \right|^2 \left| t \! \left(\omega_0\right) \phat \right|^2} \right) \wheq \!, \nonumber\\
\end{eqnarray}
reads
\begin{eqnarray}
\frac{\delta \mathcal{T}}{\delta \phat^\dagger} & = & \int \! d\omega \Bigg[
\left( t^\dagger \!\left(\omega\right) t \!\left(\omega_0\right) + 
t^\dagger \!\left(\omega_0\right) t \!\left(\omega\right)  \right)
\nonumber \\ 
&  &+
 \phat
\frac{-\phat^\dagger t^\dagger \!\left(\omega_0\right) t \!\left(\omega\right) \phat}
{\left| \treq \phat \right|^2 \left| t \!\left(\omega_0\right) \phat \right|^2} 
 \\
&  &+ 
t^\dagger \!\left(\omega\right) t \!\left(\omega\right) \phat \,
\frac{\left| \phat^\dagger t^\dagger \!\left(\omega\right) t \!\left(\omega_0\right) \phat \right|^2}
{\left| \treq \phat \right|^4 \left| t \!\left(\omega_0\right) \phat \right|^2}
\nonumber \\ 
&  &+
 t^\dagger \!\left(\omega_0\right) t \!\left(\omega_0\right) \phat \, \frac{\left| \phat^\dagger t^\dagger \!\left(\omega\right) t \!\left(\omega_0\right) \phat \right|^2} {\left| \treq \phat \right|^2 \left| t \!\left(\omega_0\right) \phat \right|^4} \Bigg] \wheq \!.\nonumber \label{grad_T}
\end{eqnarray}
Please note that $| \treq \phat |^2 = \phat^\dagger t^\dagger \!\left(\omega\right) \treq \phat$ and that a derivative with respect to $\phat$ does not lead to equations linearly independent from the set of equations (\ref{grad_T}), since $\mathcal{T} \! ( \phat )$ is a real quantity.

The $n+1$-th optimization step for the vector $\phat$ that is optimized for is then calculated from the $n$-th step according to
\begin{equation}\label{optimization-step}
\phat_{n+1} = \frac{1}{N_{n+1}} \left( \phat_n - \frac{\delta \mathcal{T}}{\delta \phat^\dagger}|_{\phat_n} \Delta s \right),
\end{equation} 
with a suitably chosen stepsize $\Delta s$ and the normalization constant $N_{n+1}$ assuring $| \phat_{n+1} | = 1$. The starting guess for each super-PM optimization is the respective projection of the widest PM onto the subspaces orthogonal to each of the previously calculated super-PMs. In the first iteration step, we start directly from the widest PM.

\section{E: super-PMs as approximate eigenstates of the $\rho$-operator}
In contrast to the time delay operator $q$, the functional $\mathcal{T}$ introduced in Eq.~(3) of the main text  involves the transmission matrix for a whole spectral interval rather than for a single wavelength only. This information benefit is one of our algorithm's key ingredients allowing for super PMs. In order to explain the mechanism causing the increased bandwidths of super-PMs, we return to the previously defined linear, $\omega$-dependent $\rhoeq$ operator in Eq.~(2) of the main text. As briefly discussed there, an eigenstate of all $\rho$-operators in the spectral region of interest would feature a perfectly flat correlation function. As we will see in the following, the super-PMs, indeed, are closer to being such mutual eigenstates than random inputs or even the PMs. For that purpose, we define a measure that quantifies how close an input state is to being a simultaneous eigenvector of all $\rhoeq$ operators,
\begin{equation} \label{recurrence-function}
B \! \left( \phat \right) := \frac{1}{\bar{N}_r} \int \! d\omega \, \frac{\left\langle \left| \rhoeq - \left\langle \rhoeq \right\rangle \right|^2 \right\rangle} {\left| \left\langle \rhoeq \right\rangle \right|^2} \wheq.
\end{equation}
The enumerator of the integrand on the r.h.s. of Eq.~(\ref{recurrence-function}) is the variance, the denominator is the absolute square of the expectation value of $\rhoeq$ for a given input state $\phat$ (see section C above for the exact mathematical expressions). The constant $\bar{N}_r$ normalizes $B$ such that the average over a large number ($10^5$) of random inputs $\bar{B}_r = 1.0$. If $\phat$ is a perfect common eigenstate to all $\rhoeq$ within the spectral range for optimization, all variances and, therefore, $B$ vanishes. The average for the 20 widest PMs evaluates to $\bar{B}_{\rm PM} = 3.1 \times 10^{-4}$ which is almost 4 orders of magnitude smaller than $\bar{B}_r$. The corresponding average for the 20 measured super-PMs, however, is $\bar{B}_{\rm SPM} = 1.3 \times 10^{-4}$ which is 2.4 times smaller than for the PMs. We therefore draw the conclusion that although the PMs are already close to being a mutual eigenstate of all relevant $\rhoeq$, the super-PMs fulfill this property even better and thereby manage to have an even broader bandwidths.

\section{F: Characteristics of Super- and Anti-PMs in the MMF}

\begin{figure}[b]
  \begin{center}
    \includegraphics[angle=0, scale=1.0, width=\columnwidth]{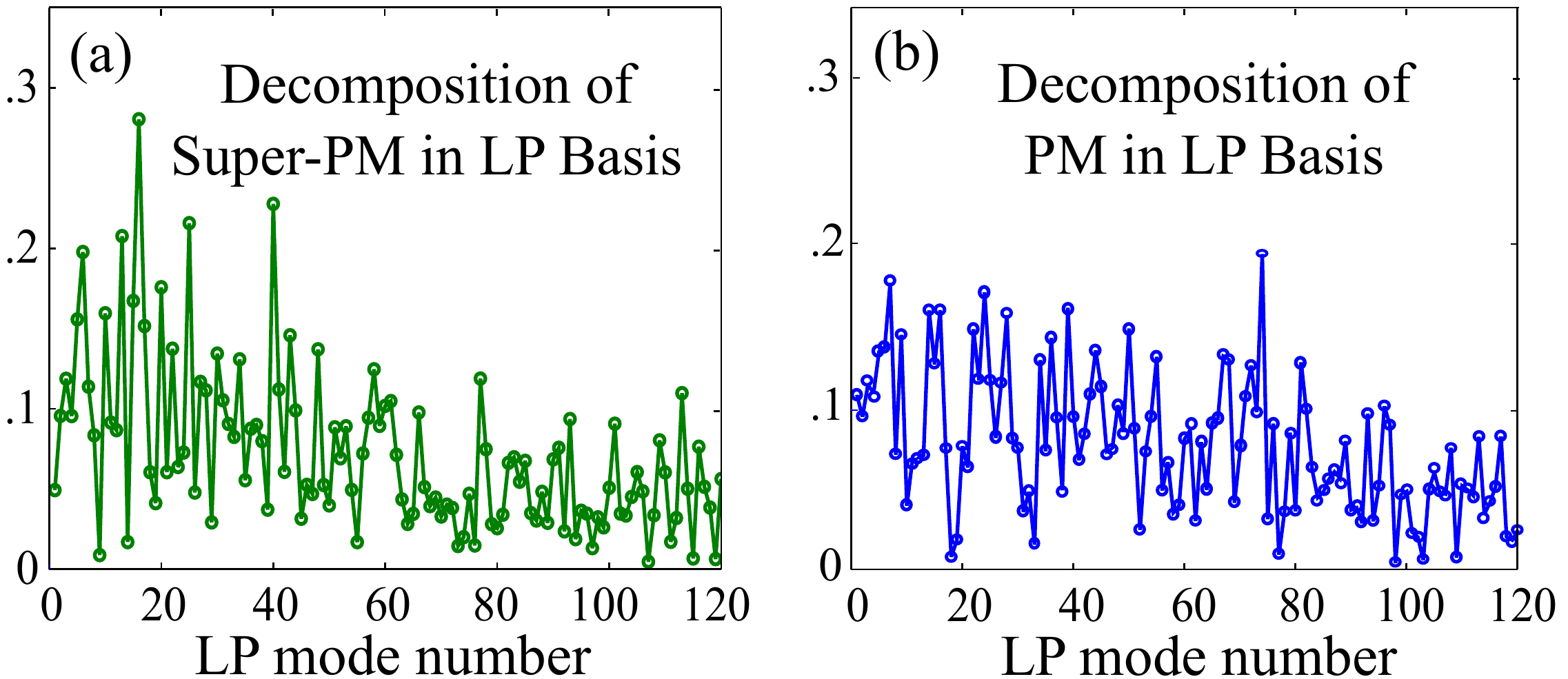}
    \caption{ Decomposition of an experimentally measured super-PM (left) and a PM (right) by LP modes of the MMF. The LP modes are numbered by their propagation constant from large to small. No specific enhancements or features are discernible, wherefore we use instead the decomposition of super- and anti-PMs in the PM basis (see main text).} 
    \label{fig:SM1}
  \end{center}
\end{figure}

First we decompose the experimentally measured super-PMs and the conventional PMs in the LP mode basis. In Fig.~\ref{fig:SM1} we show as an example the decomposition of both a super-PM [Fig.~\ref{fig:SM1}(a)] and a PM [Fig.~\ref{fig:SM1}(b)]: in both cases the decomposition shows no discernible structure.     
In contrast, the decomposition of the super-PMs in the PM basis [see Fig.~3(a) in the main text] shows that super-PMs are composed of PMs with similar delay times, shedding light on the formation of super-PMs. 
\nopagebreak

\begin{figure}[h]
\nopagebreak
  \begin{center}
    \includegraphics[angle=0, scale=1.0, width=\columnwidth]{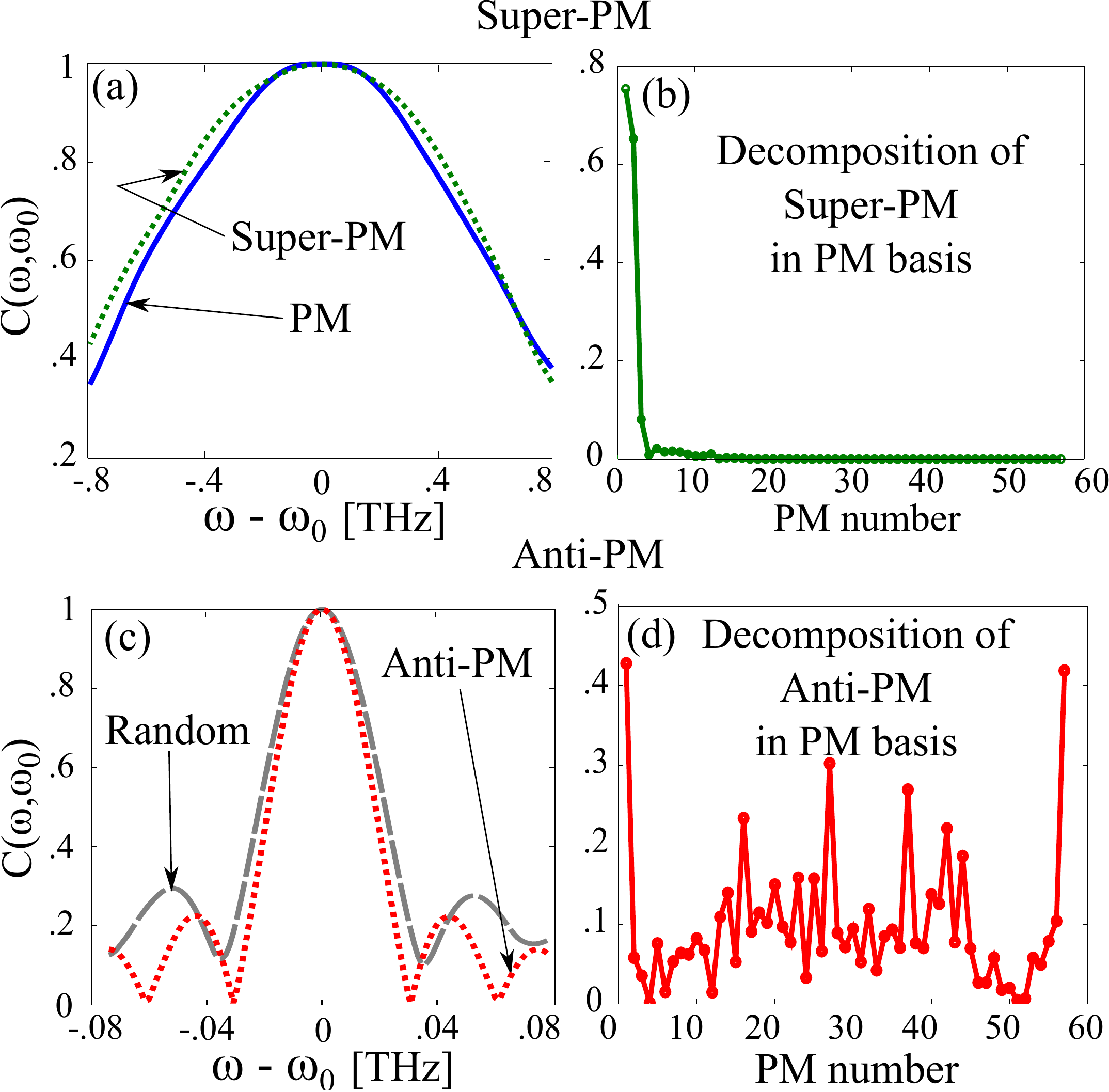}
    \caption{ Super-PM and anti-PM in the weak mode coupling regime without MDL. (a) Spectral correlation function of the widest super-PM and the widest PM. They have similar spectral correlations. (b) Decomposition of the widest super-PM in the basis of PMs, which are numbered from short to long delay time. The super-PM is mainly composed of two PMs with the shortest delay times and the widest spectral correlations. (c) The anti-PM yields a similar spectral correlation as the random input. (d) Decomposition of the anti-PM in the PM basis. PMs with intermediate delay times are involved to reduce the side peaks of spectral correlation function.  } 
    \label{fig:SM2}
  \end{center}
\end{figure}

In the weak mode coupling regime, our optimization scheme for constructing super-PMs or anti-PMs does not yield a significantly different spectral correlation width as compared to conventional PMs or random inputs [Fig.~\ref{fig:SM2}(a) and (c)]. We numerically studied super-PMs and anti-PMs in the weak mode coupling regime using the same numerical fiber model as the one described in the main text.   For the super-PM in the weak mode coupling regime, since the delay time difference between adjacent PMs is increased, the super-PM is composed of fewer PMs as seen in Fig.~\ref{fig:SM2}(b). The spectral correlation width of the super-PM, therefore, is almost identical to the bandwidth of the dominant PM that constitutes the super-PM [Fig.~\ref{fig:SM2}(a)]. 

\begin{figure*}
	\includegraphics[angle=0, scale=1.0, width=2\columnwidth]{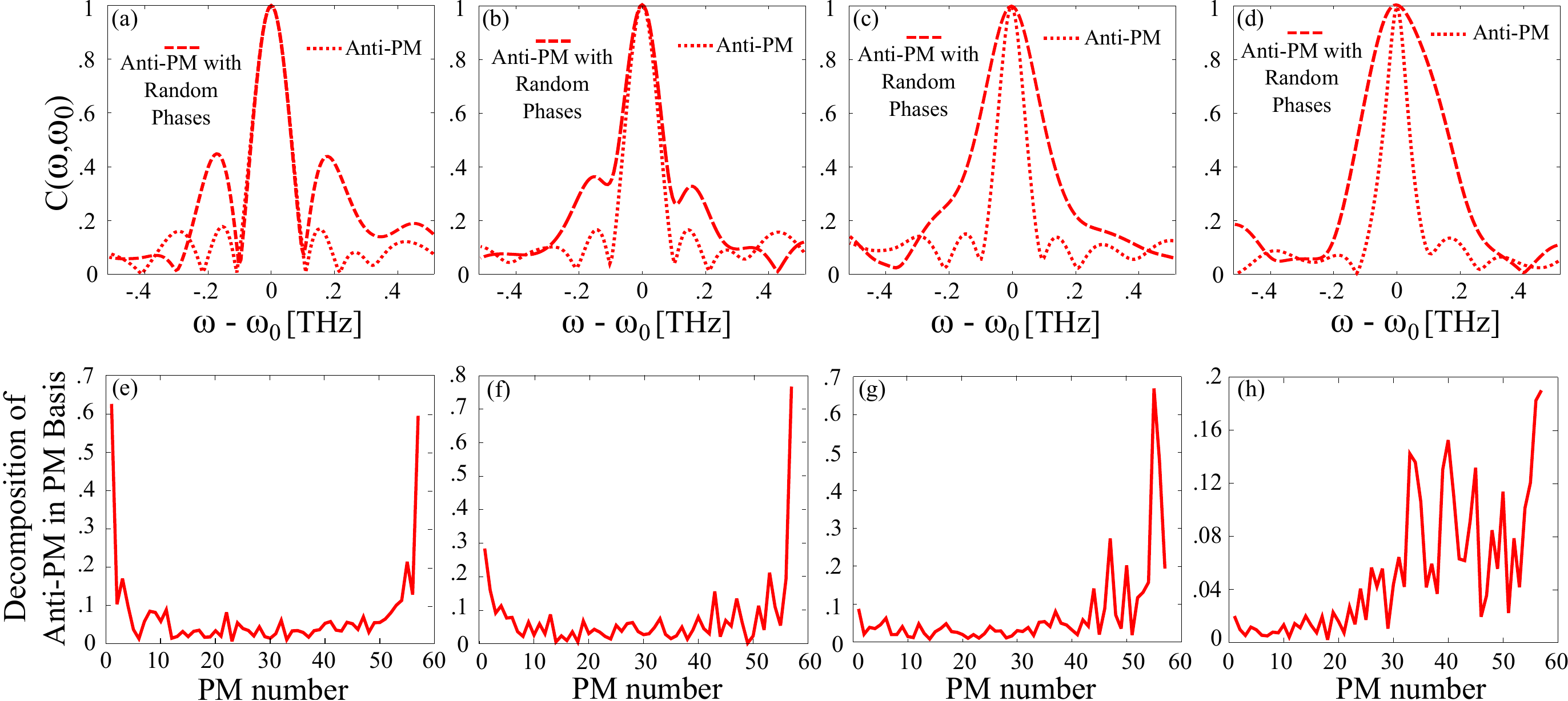}
	\caption{ Effect of MDL on anti-PMs. (a)-(d) Spectral correlation function of the narrowest anti-PM (red dotted line) in a multimode waveguide with MDL = 0 dB, -3.5 dB, -7.5 dB and -20 dB correspondingly. The red dashed line is obtained by randomizing the phases of constituent PMs in the anti-PM. The phase randomization significantly increases the spectral correlation width at large MDL, where the formation of anti-PM relies more on the interference of constituent PMs.  (e-h) Decomposition of the anti-PMs in (a-d) in the basis of PMs, which are numbered from short to long delay times. As the MDL increases, the PM with long delay time gains weight to compensate for the stronger loss.} 
	\label{fig:SM3}
\end{figure*}

For the anti-PM in the weak mode coupling regime, the PM-decomposition includes more PMs with intermediate delay times as compared to the case of strong mode coupling. The reason is that in the weak mode coupling regime, the fastest PM and the slowest PM have very different delay times. The beating between the two PMs is appreciable and thereby will result in the revival of spectral correlation. To reduce the beating and revival, additional PMs with intermediate delay times are included as seen in Fig.~\ref{fig:SM2}(d). Meanwhile, the inclusion of many PMs with medium delay times makes the anti-PM less distinctive from random inputs. Thus the spectral correlation functions of the anti-PM and random inputs become similar [Fig.~\ref{fig:SM2}(c)].

In a next step we also studied the effect of MDL on anti-PMs numerically. The numerical model is a 200 $\mu m$ wide waveguide with 0.22 numerical aperture, the same as the model in the main text. The MDL is quantified by $(\tau_{max} - \tau_{min})/{\tau_{max}} $, where $\tau_{max}$ is the highest transmission value and $\tau_{min}$ is the lowest transmission value. For simplicity, we assume a spatially uniform absorption coefficient across the fiber, and by varying the absorption coefficient, we change the MDL. Fig.~\ref{fig:SM3}(a-d) show the spectral correlation function of the narrowest anti-PM (black solid line) for MDL = 0 dB, -3.5 dB, -7.5 dB and -20 dB, respectively. We se that the anti-PM has narrower spectral correlation width when the MDL is larger. Fig.~\ref{fig:SM3}(e-h) show the decomposition of the anti-PM in the PM basis, which changes with increasing MDL. The PMs are numbered by their delay times from short to long. In the absence of MDL [Fig.~\ref{fig:SM3}(e)], the anti-PM is mainly composed of PMs with extreme delay times. As the MDL increases [Fig.~\ref{fig:SM3}(f)-(h)], the contribution from PMs with long delay time increases, to compensate the greater loss they experience. Meanwhile, more PMs with intermediate delay times start to participate in the formation of anti-PM. We also randomized the relative phases of the constituent PMs in the anti-PM, and recalculated the spectral correlation function [red dashed line in Fig.~\ref{fig:SM3}(a-d)]. Without MDL, the main peak of the spectral correlation function remains almost the same as that of the anti-PM [Fig.~\ref{fig:SM3}(a)]. With an increase of the MDL, the main spectral correlation becomes broader, as the relative phases between constituent PMs become more important [Fig.~\ref{fig:SM3}(b)-(d)]. Thus the formation of anti-PMs relies more on the interference of PMs when the MDL is increased.

\bibliography{OSPM}